\documentclass[aps,pra,twocolumn,amsmath,amssymb,amsfonts,superscriptaddress,floatfix,footinbib]{revtex4-1}
\usepackage{epsfig} \usepackage{graphicx}
\usepackage{color}
\usepackage{multirow}
\usepackage{wasysym}
\usepackage[pdftex,bookmarks]{hyperref}
\usepackage[all]{hypcap}

\newcommand{\bgar}{\begin{eqnarray}}
\newcommand{\enar}{\end{eqnarray}} 
 
 \newcommand{\be}{\begin{equation}}
\newcommand{\ee}{\end{equation}}  
 \def\mincirc{\lower
  3pt\hbox{$\buildrel<\over{\hbox{$\mathchar"218$}}$}}

\begin{document}

\title{\bf Do general relativistic effects limit experiments to test the universality of free fall and the weak equivalence principle? \\ }

 \date{\today}

\author{Anna M. Nobili}
\affiliation{
Dept. of Physics ``E. Fermi'', University of Pisa, Largo B. Pontecorvo 3,
56127 Pisa, Italy}
\affiliation{INFN-Istituto Nazionale di Fisica Nucleare, Sezione di Pisa, Largo B. Pontecorvo 3,  56127 Pisa, Italy}

\begin{abstract}

The Universality of Free Fall and the Weak Equivalence Principle, which are at the basis of General Relativity, have been confirmed  to $1$ part in $10^{13}$.  Space experiments with macroscopic test masses of different composition orbiting the Earth inside a low altitude satellite aim at improving this precision by two orders of magnitude (with the Microscope satellite, launched on 25 April 2016) and up to four orders of magnitude  (with the ``Galileo Galilei'' - GG satellite). At such a high precision many tiny effects must be taken into account in order to be ruled out as the source of a spurious violation signal. In this work we investigate the general relativistic effects, including those which involve the rotation of both the Earth and the test masses,  and show that they are by far too small to be considered  even in the most challenging experiment.\footnote{Paper to appear  on Physical Review D}

\end{abstract}

\maketitle

\section{Introduction}\label{Sec:Introduction}

The Universality of Free Fall (UFF), also known as the Weak Equivalence Principle (WEP) refers to the fact  that  in a gravitational field all bodies fall with the same acceleration regardless of their mass and composition (see e.g.~\cite{NobiliAJP2013}).  As stated by Einstein in 1916~\cite{Einstein1916GTR} the General theory of Relativity (GR) requires this fact to hold.

The best experimental tests so far involve artificial  proof masses suspended on rotating torsion balances~\cite{Adel99etasun,Adel2008,TorsionBalancesFocusIssue} or  celestial bodies (Earth and Moon, through  Lunar Laser Ranging~\cite{LLR2004,LLRfocusIssueJPL,LLRfocusIssueMueller}). They show no anomalous differential acceleration between the falling bodies --in the fields of the Earth and the Sun-- to about $10^{-13}$. 

Considerable improvements are expected in the field of the Earth by flying macroscopic proof masses inside a low altitude spacecraft. Microscope~\cite{MicroscopeFocusIssue},  launched  in April 2016,  aims at $10^{-15}$, and ``Galileo Galilei'' (GG)~\cite{GGfocusIssue} aims at $10^{-17}$.  As for cold atom tests, at present they have reached a few  $10^{-8}$~\cite{UFFcoldAtomsCinesi2015}  and are not expected to compete with high-precision space tests based on macroscopic bodies~\cite{ColdAtoms2016}.

Despite their high precision goals, and consequent very small effects to be measured, the GG and Microscope experiments are designed and investigated within Newtonian mechanics, in the assumption that general relativistic effects  are too small to compete with the sought for violation signal. 

During a recent competitive selection of  proposals shortlisted for  the  medium size mission M4 of the European Space Agency (ESA)  this assumption has been questioned by SARP, the  ``Science Assessment Review Panel'' appointed by  ESA   to evaluate the GG proposal. From their Report on GG we read~\cite{SARPM42015}: 

\begin{quotation}
 ``The breakdown of the WEP is sought in the frame work of the response of test matter to terrestrial Newtonian  gravitation. The  source  of  terrestrial  Newtonian  gravitation  is  independent  of  the  Earth's  (non uniform)  rotation.  Furthermore,  the  test  cylinders  in  the  proposed  experiment  are  spinning. In General Relativity the gravitational field of a spinning source depends on its spin. Also the  mass  centroid  motion  of  extended  spinning  test  matter  in  an  external  gravitational  field  may  depend  on  its  spin  and  still  be  geodesic  (independent  of  inertial  mass)  when  its  spin  is  zero.  The  estimates,  based  on  General  Relativity,  of  the  effect  of  the  Earth's  rotation  on  the  motion  of  each  spinning cylinder or the laser interferometer and their relevance to the interpretation of any non null  signal  at  the  expected  level  of  accuracy  have  not  been  sufficiently  explained  to  the  satisfaction  of  the SARP.''
 \end{quotation}

Experimental evidence of a  violation of UFF-WEP  would require either a modification/augmentation of  GR or the existence of a new composition dependent force of nature. Either way, it would make a revolution in physics. On the other hand,  a null result to a very high precision will be a landmark for any attempt at overcoming the current physics impasse. The situation is reminiscent of that at the end of the 19th century, when Michelson and Morley tested by very precise light interferometry the propagation of the newly discovered electromagnetic waves through the ether and proved that the ether does not exist\cite{MM1887}; a very precise null result which in 1905 led to the theory later named Special Relativity  (Michelson was awarded the Nobel prize in 1907).

By reaching its goal  Microscope  will improve the current best tests of UFF-WEP by $2$ orders of magnitude, to $10^{-15}$. Should the possibility of a non zero violation signal emerge from Microscope data, it will call for urgent checking and  more precise space experiments such as GG (which aims at $10^{-17}$) might become of interest to space agencies. It is therefore important and timely to firmly establish the role of general relativistic effects in high precision satellite tests of UFF-WEP.

Deviations from Newtonian predictions in the motion of orbiting bodies have been investigated since the birth of General Relativity,  in the hope  to provide observational evidence for the new theory. 

In November 1915 Einstein presented to the Prussian Academy of Sciences in Berlin his results on the ``Explanation of the perihelion motion of Mercury from the General Theory of Relativity''~\cite{Einstein1915Mercury}, and one week later ``The field equations of gravitation''~\cite{Einstein1915}. A year later De Sitter published: ``Planetary motion and the motion of the moon according to Einstein's theory''~\cite{DeSitter1916}.

Einstein's  Mercury paper and De Sitter's work mentioned above deal with non rotating masses. However, all celestial bodies rotate. According to Newton, the gravitational field of a celestial body does not depend on whether the body does rotate or not. Not so in General Relativity.  In 1918 Thirring and Lense~\cite{Thirring1918,LenseThirring1918} calculated the secular effects  of the rotation of the central body on the orbits of planets and moons (see the English translation and comments on the Thirring and Lense papers by~\cite{ThirringLensePapers}).

Since then, considerable theoretical work has been carried out to include also the rotation of the secondary body, leading to the so-called MPD  (Mathisson, Papapetrou, Dixon) equations~\cite{Mathisson1937,Papapetrou1951,Dixon1964}. The whole subject was revisited in the early 1960s~\cite{Schiff1960A,Schiff1960B,Schiff1961}  with the proposal to fly  a high precision gyroscope in low Earth orbit in order to measure general relativistic effects (the GP-B mission, launched in 2004~\cite{GPB2011}) and in the 1970s became of primary importance  for understanding binary systems made by very compact  rapidly rotating stars~\cite{BOC1970,BOC1975}. 

Similarly to the modern torsion balances used for testing UFF-WEP, the Microscope and GG satellites are designed to rotate in order to up-convert the target signal to higher frequency where important noise sources are known to be smaller than they are at lower frequencies~\cite{Adel2009,ThermalNoise2011}. 

We  compute the general relativistic effects in these experiments by referring to the literature available in which the spin angular momentum of both the primary and secondary body are taken into account. We  refer in particular to the work of Barker and O'Connel~\cite{BOC1970,BOC1975} (in checking the equation numbers  quoted please note that they refer to the primary and secondary body with the numbers `2' and  `1' respectively while in this work we do the opposite).  

This paper is organized as follows. In Sec.\,\ref{Sec:GReffectsDifferantialAcceleration} we estimate the differential accelerations between the test masses as predicted by GR, in absence of spin as well as in the presence of a spinning Earth and of spinning test cylinders. We also recall the Newtonian differential effects due to the quadrupole mass moments of the  interacting bodies. Sec.\,\ref{Sec:GReffectsOrbits} deals with differential  precession of the orbits  due to General Relativity (and Newtonian dynamics) and relates them to the  differential accelerations which give rise to them and might compete with the violation signal.  In Sec. \,\ref{Sec:GReffectsSpnAxes} the general relativistic and Newtonian effects on the spin axes of the test cylinders are estimated. In Sec.\,\ref{Sec:LaserGauge} we discuss the effects of rotation on the laser interferometry readout of GG.  The conclusions, that in all cases general relativistic effects are  negligible by and large,  are drawn in Sec.\,\ref{Sec:Conclusions}.

\section{General relativistic accelerations competing with a violation signal}
\label{Sec:GReffectsDifferantialAcceleration}

Tests of the Universality of Free Fall are quantified  by  the fractional differential acceleration
\begin{equation}\label{eq:eta}
\eta=\frac{\Delta a}{a}
\end{equation}
between two test masses of different composition  as they fall in the gravitational field of a source body  with the  average  acceleration $a$ (``driving signal''). The physical observable  is the differential acceleration $\Delta a$ of the falling masses relative to each other, pointing to the center of mass of the source body.  

For test masses orbiting the Earth inside a low altitude, low eccentricity, sun-synchronous satellite such as Microscope or GG a violation signal ($\Delta a\neq0$) driven by the Earth would have the orbital frequency while the driving signal $a$  at the denominator  is the gravitational acceleration $g(h)$ caused by the Earth at the satellite altitude $h$;  $h\simeq630\,\rm km$ for GG  (with $g(h)\simeq8.1\,\rm ms^{-2}$) and slightly higher for Microscope. 

Up-convertion of the signal to higher frequency (the higher, the better) is regarded by all experimentalists as a crucial asset because  thermal and electronic noise are  lower at higher frequency (the higher the better). This fact has been demonstrated by the rotating torsion balances, which have been able to reach the thermal noise limit expected at their rotation rate (see~\cite{Adel2009}, Fig.\,20). 

For this reason both Microscope and GG are designed to rotate, though the way they accomplish it is different due to the different  experiment design. 

The test masses are (nominally) concentric hollow cylinders in both cases. 

A very good coincidence of the centers of mass is crucial because of  a  major Newtonian effect caused by  off-centering.  Because of the non uniformity of the gravitational force there is a tidal differential acceleration from the Earth (non zero gravity gradient); it acts at twice the orbital frequency, but there is also a smaller tidal effect, proportional to the eccentricity of the orbit, which acts at the orbital frequency and therefore competes directly  with the sought for violation signal. This is a major  limiting factor to space tests of UFF, which cannot be totally eliminated because it is impossible to  inject the satellite in an exactly circular orbit.   Indeed, for  high precision tests the centers of mass of the test cylinders must be centered on one another  far  better than  it is typically  achieved at the time of launch by construction and mounting. 

In Microscope the test cylinders are required  to be concentric within $20\,\rm \mu m$ at launch, and no further adjustment is performed in space. 
Microscope scientists plan to use  the tidal effect at twice the orbit frequency  within  \textit{a posteriori} data analysis in order to reduce --from the measurement of this effect-- the  unknown level of off-centering between the test cylinders to within $0.1\,\rm \mu m$, i.e. a factor $200$ better than achieved via hardware  at launch (see~\cite{MicroscopeFocusIssue}, p.\,4). A  recent press release  by the French space agency  reports  measured offsets of $25\,\mu\rm m$ and $33\,\mu\rm m$\,\cite{CNESpressRelease160929} (the two values are likely to refer to the two accelerometers carried by the Microscope satellite, each one with two test cylinders).
 
In Microscope each cylinder is constrained to move along its symmetry  axis (sensitive axis): weak electrostatic coupling along the axis; one degree of freedom; effect of violation signal maximized when the symmetry axis points to the center of mass of the Earth.  Any differential effect (including the  violation signal, if any) would displace the centers of mass of the two cylinders relative to each other. An active control loop ensures that they remain centered: the control force itself contains the violation signal along with  all classical (and GR) differential effects.
In order to up-convert the frequency of the signal  to higher frequency the sensitive/symmetry axis must rotate relative to the satellite-to-Earth direction, hence rotation must  occur around an axis perpendicular to the symmetry axis of the cylinders. However, it is known in classical mechanics that a rotating axisymmetric rigid body is stable to small perturbations only if rotation occurs around the axis whose principal moment of inertia is distinct from the other two. The rotation mode of Microscope is slow and actively controlled, up to a maximum rate of about $\frac{1}{900}\,\rm Hz$, with roughly a factor of $7$  up-conversion from  the orbital/signal frequency in absence of rotation (\cite{MicroscopeFocusIssue},\cite{TouboulMicroscopeColl2015}).

In GG the cylinders are allowed to move in the plane perpendicular to the symmetry axis (sensitive plane), where they respond to any differential acceleration: weak mechanical coupling in the plane; two degrees of freedom; effect of violation signal maximized when the symmetry axis is perpendicular to the orbit plane.  The violation signal would displace the test cylinders to a new equilibrium position and the displacement is measured by a readout laser gauge.
In accordance with the cylindrical symmetry of the system rotation  occurs around the symmetry axis, and it is stable, hence passive attitude stabilization of the satellite is ensured (no active attitude control needed). At the same time the signal is up-converted from the orbital frequency of $1.7\cdot10^{-4}\,\rm Hz$ to the much higher rotation frequency of  $1\,\rm Hz$ (with an up-conversion factor of almost $6000$)  where electronic and thermal noise are much lower~\cite{ThermalNoise2011,IntegrationTimePRD2014}.  
The rotation frequency is provided at the start of the mission and maintained by angular momentum  conservation. Hence the whole satellite spins  with no need of  motor and bearings, which are a well known major source of noise for all rotating  experiments in ground laboratories.  

A rotating conductor in the magnetic field of the Earth is known to slow down because of energy dissipation due to eddy currents induced in the conductor by a component of the magnetic field perpendicular to the spin axis. In GG the largest such effect will take place in the $Be$ test cylinder (the outer shell of the spacecraft will be manufactured in  carbon fiber, and the inner  $Ti$  test cylinder  has smaller size and lower conductivity). This effect has been calculated in\,\cite{IntegrationTimePRD2014}, Sec.\,IIIB, and found to be extremely small: the $Q$ factor of spin energy dissipation is $1.4\cdot10^{10}$, which means that in the total 1 year duration of the mission the spin frequency will decrease by about $1\%$. This value has been obtained in the presence of a magnetic field of the Earth reduced by a factor $150$ by means of a $\mu$-metal shield, and under worst case assumptions. Any residual differential rotation between the outer shell of the GG spacecraft and its inner parts will be sensed and compensated, if needed,  by means of the cold gas thrusters in charge of compensating for non gravitational forces (mostly drag from residual atmosphere).

Since the  cylinders are suspended and coupled very weakly (taking advantage of the absence of weight in orbit), the frequencies of their  normal modes  are much lower than the spin frequency. This is a dynamical regime known as  `super-critical rotation' (spin speed above the normal mode/critical speed), which ensures `self-centering' better than achieved via construction and mounting  by as much as the ratio of the spin-to-normal mode frequency squared. It is well known that such self-centering by physical laws  requires two degrees of freedom: see Ch.\,6 of Den Hartog textbook~\cite{DenHartog}, particularly Eq.\,(6.2) for self-centerimg and Fig.\,6.4 for evidence of rotation instability for systems with  $1$ degree of freedom. It cannot therefore be exploited in Microscope. 

It is also well known that the presence of non zero internal damping in the rotating system (rotating damping) gives rise to a slowly growing `whirling' motion at the normal mode frequency: the smaller the damping, the weaker the instability, the slower its growth, the smaller the fraction of the suspension force  which is required to damp it (see~\cite{DenHartog} Sec.\,7.4; \cite{Genta} Sec.\,4.5, and~\cite{Nobili1999}). In GG  self-centering by physical laws  is ensured  at  a few tens of picometers, and whirl is damped by capacitance sensors/actuators so as not to exceed a separation level between the centers of mass of $1.7\,\rm nm$; a noise  well within the reach of capacitance bridges. Whirl damping is off during science data taking so that the test masses are totally passive save for the laser light of the interferometer in charge of reading their differential displacements.

Only differential accelerations between the test masses do compete with the target violation signal. Accelerations caused by the primary body (the Earth) and ascribed to GR have a specific dependence  on the orbiting distance  of each test mass; its first order differential effect is  linear with the offset $\Delta r$ of the test masses in the direction to the the center of mass of the Earth, which may mimic a violation signal.

All effects  predicted by GR on the test masses contain the very small dimensionless parameter:
\begin{equation}\label{eq:GRparameter}
\epsilon=\frac{GM}{c^{2}r}\simeq6.3\cdot10^{-10}
\end{equation}
where  $G$ is the universal constant of gravity,  $c$ the speed of light, $M$  the mass of the Earth and $r$ the orbital distance of the test body around it. The small parameter $\epsilon$  is the ratio between the Schwarzschild radius of the Earth $GM/c^{2}=4.4\,\rm mm$ and the satellite orbital distance $r=R+h\simeq7\cdot10^{6}\,\rm m$.

The largest GR acceleration on each test mass, $a_{E}$, was computed by Einstein in 1915~\cite{Einstein1915Mercury} for non rotating interacting bodies (see also~\cite{BOC1970} Eq.\,(65a)). The second largest one, $a_{S_{1}}$, is due to the spin of the primary body (spin-orbit interaction; see~\cite{BOC1970} Eq.\,(65c)) while the smallest one, $a_{S_{1}S_{2}}$,  is due to the fact that both the primary body and the test mass are spinning (spin-spin interaction; see\cite{BOC1970} Eq.\,(65d)). Their respective orders of magnitude are: 
\begin{equation}\label{eq:EinsteinAcceleration}
a_{E}\simeq4\epsilon g(h)\simeq2\cdot10^{-8}\,\rm ms^{-2}
\end{equation}
\begin{equation}\label{eq:SpinOrbitAcceleration}
a_{S_{1}}\simeq6\epsilon\frac{S_{1}}{M}\frac{v}{r^{2}}\simeq5.7\cdot10^{-10}\,\rm ms^{-2}
\end{equation}
($S_{1}\simeq0.33MR^{2}\omega_{1}$ 
 is the spin angular momentum of the Earth  with angular velocity  $\omega_{1}\simeq7.3\cdot10^{-5}\,\rm rad\,s^{-1}$, hence $\frac{S_{1}}{M}\simeq9.8\cdot10^{8}\,\rm m^{2}s^{-1}$; $v\simeq7.5\cdot10^{3}\,\rm m\,s^{-1}$ is the orbital velocity of the satellite  at distance $r\simeq7\cdot10^{6}\,\rm m$)
 and:
\begin{equation}\label{eq:SpinSpinAcceleration}
a_{S_{1}S_{2}}\simeq3\epsilon\frac{S_{1}}{M}\frac{S_{2}}{m}\frac{1}{r^{3}}
\simeq4.8\cdot10^{-22}\,\rm ms^{-2}
\end{equation}
with $S_{2}$ the spin angular momentum of a test body of  mass $m$. For  the hollow test cylinders of GG spinning around the symmetry axis with angular velocity $\omega_{2}=2\pi\,\rm rad\,s^{-1}$, inner radius $a$, outer radius $b$, height $H$, it is  $S_{2}=\frac{1}{2}m(a^{2}+b^{2})\omega_{2}$ hence, in the worst case (largest value of $S_{2}$) of the outer cylinder ($a\simeq10.5\,\rm cm$, $b\simeq13\,\rm cm$, $H\simeq28.6\,\rm cm$), it is  $\frac{S_{2}}{m}\simeq0.088\,\rm m^{2}\,s^{-1}$.

What matters in UFF/WEP tests is the differential acceleration between the test masses. In the case of the GR effects    (\ref{eq:EinsteinAcceleration}) and (\ref{eq:SpinOrbitAcceleration})  a differential acceleration arises because of   a non zero offset $\Delta r$ between the  centers of mass of the test cylinders. In the case of GG, with $\Delta r\simeq 1.7\,\rm nm$ as reported above, we have: 
\begin{equation}\label{eq:DifferentialEinsteinAcceleration}
\Delta a_{E}\simeq3a_{E}\frac{\Delta r}{r}\simeq1.5\cdot10^{-23}\,\rm ms^{-2}
\end{equation}
\begin{equation}\label{eq:DifferentialSpinOrbitAcceleration}
\Delta a_{S_{1}}\simeq\frac{7}{2}a_{S_{1}}\frac{\Delta r}{r}\simeq4.8\cdot10^{-25}\,\rm ms^{-2}\ \ .
\end{equation}
Instead, the acceleration  (\ref{eq:SpinSpinAcceleration}) depends on the geometrical properties of the test cylinders, which are necessarily different because they have been designed to be one inside the other,  yielding a differential acceleration  larger than the one caused by off-centering. A worst case assumption is for the differential acceleration to be of same order as the the acceleration itself:
\begin{equation}\label{eq:DifferentialSpinSpinAcceleretaion}
\Delta a_{S_{1}S_{2}}\simeq a_{S_{1}S_{2}}\simeq3\cdot10^{-22}\,\rm ms^{-2}  \ \ \ .
\end{equation}

Should the differential accelerations~(\ref{eq:DifferentialEinsteinAcceleration})-(\ref{eq:DifferentialSpinSpinAcceleretaion})  not be identified as due to General Relativity, they might be misinterpreted as a violation of UFF/WEP  at the corresponding (spurious) levels:
\begin{equation}\label{eq:EtaE}
\eta_{E}=\frac{\Delta a_{E}}{g(h)}\simeq1.8\cdot10^{-24}
\end{equation}
\begin{equation}\label{eq:EtaS1}
\eta_{S_{1}}=\frac{\Delta a_{S_{1}}}{g(h)}\simeq6\cdot10^{-26}
\end{equation}
\begin{equation}\label{eq:EtaS1S2}
\eta_{S_{1}S_{2}}=\frac{\Delta a_{S_{1}S_{2}}}{g(h)}\simeq3.7\cdot10^{-23}
\end{equation}
showing that even the largest one, caused by the spin-spin interaction, is smaller that the GG target $\eta_{GG}=10^{-17}$  by more than $5$ orders of magnitude. There is therefore no need to investigate the specific signature (frequency, phase, dependence on the orbital parameters) and the exact values of these effects which  come into play in the framework of General Relativity. 

In the case of Microscope the spin-spin effect (\ref{eq:EtaS1S2})  is even less relevant than it is  for GG because of  a smaller value of $S_{2}/m$ (mostly because of a slower spin rate by almost $3$ orders of mangnitude) and also because of the lower precision target $\eta_{microscope}=10^{-15}$ of the mission. As for (\ref{eq:EtaE}) and (\ref{eq:EtaS1}) they are about $60$ times larger for Microscope (assuming  $\Delta r\simeq0.1\,\rm\mu m$ as reconstructed a posteriori) hence their ratio to the mission target   is about a factor of two smaller than the corresponding one for GG.

It is worth noticing that, as  expected, the general relativistic effects considered above are much smaller than the Newtonian ones due to the non-zero quadrupole mass moments of the Earth and the test masses. 

The quadrupole mass moment of the Earth $J_{2}^{(1)}\simeq10^{-3}$ gives rise to an additional acceleration on each test mass (see e.g.~\cite{BOC1970}, Eq.\,(65e))
\begin{equation}\label{eq:Q1acceleration}
a_{Q_{1}}\simeq\frac{3}{2}g(h)J_{2}^{(1)}\bigg(\frac{R}{r}\bigg)^{2}\simeq10^{-2}\,\rm ms^{-2} \ \  .
\end{equation}
If the centers of mass of the test bodies are well centered on one another its differential value is below the target. For GG we have:
\begin{equation}\label{eq:Q1DifferentialAcceleration}
\Delta a_{Q_{1}}\simeq4a_{Q_{1}}\frac{\Delta r}{r}\simeq10^{-17}\,\rm ms^{-2}
\end{equation}
hence
\begin{equation}\label{eq:etaQ1}
\eta_{Q_{1}}=\frac{\Delta a_{Q_{1}}}{g(h)}\simeq1.3\cdot10^{-18}
\end{equation}
which is a factor $8$ smaller than the mission target  $\eta_{GG}=10^{-17}$.  

If each test cylinder has a non zero  quadrupole mass moment  because its principal  moments of inertia are not all equal, the Earth's monopole does couple with it yielding an additional (Newtonian) acceleration which has been known as a major limitation to tests of UFF/WEP aiming at very high precision~\cite{STEPESA1993,STEPESA1996}. 
In the specific configuration of the GG experiment, with principal moments of inertia $I_{z}$ relative to the spin/symmetry axis, $I_{x}=I_{y}$ relative to the cartesian axes in the plane perpendicular to it and a non zero value of  the ratio $\frac{\Delta I}{I_{x}}=\frac{I_{z}-I_{x}}{I_{x}}$,  this effect has been calculated to be of the order (see~\cite{GGphaseA1998}, Sec.\,2.2.5): 
\begin{equation}\label{eq:Q2acceleration}
a_{Q_{2}}\simeq\frac{3}{8}g(h)\frac{\Delta I}{I_{x}}\frac{r_{Q}^{2}}{r^{2}}
\end{equation}
with  $r_{Q}^{2}=a^{2}+b^{2}+\frac{H^{2}}{3}$. The corresponding differential acceleration is dominated by the different value for   the two test cylinders of the factor  $\frac{\Delta I}{I_{x}}r_{Q}^{2}$. In GG they have been designed   so as to make this effect much smaller than the target signal:
\begin{equation}\label{eq:Q2DifferentialAcceleration}
\Delta a_{Q_{2}}\simeq5.6\cdot10^{-18}\,\rm ms^{-2}
\end{equation}
yielding a (spurious) violation at the level:
\begin{equation}\label{eq:etaQ2}
\eta_{Q_{2}}=\frac{\Delta a_{Q_{2}}}{g(h)}\simeq6.9\cdot10^{-19}
\end{equation}
which is about a factor $14$ below the target. Note that this result has been obtained with a fractional difference in the moments of inertia for each test cylinder of the order of $0.01$, which is not a demanding requirement at all. 

Should GG aim at $10^{-18}$ with the same level of  centering of the test masses, both effects (\ref{eq:Q1DifferentialAcceleration}) and (\ref{eq:Q2DifferentialAcceleration}) would be close to the target signal. 
 However, their  signature is different from that of the signal and known exactly from celestial mechanics (in the case of (\ref{eq:Q1DifferentialAcceleration}) the value of $J_{2}$ of the Earth is well determined in satellite geodesy). Hence they can be separated from the signal by means of various measurements,  each one to the target precision, in different dynamical conditions (e.g. different angles between the spin axis and the normal to the orbit plane). Many such measurements are possible because of the short integration time required by GG~\cite{IntegrationTimePRD2014}.

In the case of Microscope a correct estimate of the effect (\ref{eq:Q2acceleration}) should be calculated  taking into account the specific geometry and mass distribution of the test bodies. However, for the  target $\eta_{{Microscope}}=10^{-15}$ this effect is not a matter of concern, and it has never been listed in the error budget of the mission.

\section{General relativistic effects on the orbits of the test masses}
\label{Sec:GReffectsOrbits}

In a 2-body problem the secular effects due to General Relativity on the semimajor axis of the orbit are zero. The orbital  angular momentum vector (perpendicular to the orbit plane) and the Lenz vector (a vector pointing  to the pericenter of the orbit whose modulus is the orbital eccentricity) which within Newtonian gravity  are both fixed in inertial space,  within GR are subjected to a secular precession with the same angular velocity.  Since the precession velocity is the same, the orbit precesses as a whole (see~\cite{BOC1970}, Eqs. (73)-(74)).

All GR contributions to  orbit  precession are  proportional to the $\epsilon$  parameter (\ref{eq:GRparameter}). There is  a contribution independent of rotation, a contribution due to the rotation of the primary body and a contribution due to the rotation of both the primary and secondary body. We have: 
\begin{equation}\label{eq:EinsteinOrbitPrecession}
\Omega^{orbit}_{E}\simeq3\epsilon n\simeq2\cdot10^{-12}\,\rm rad\,s^{-1}
\end{equation}
with $n$ the mean orbital angular velocity of the satellite (see~\cite{BOC1970}, Eq.\,(76a)),
\begin{equation}\label{eq:SpinOrbitPrecession}
\Omega^{orbit}_{S_{1}}\simeq0.66\epsilon\bigg(\frac{R}{r}\bigg)^{2}\omega_{1}\simeq2.5\cdot10^{-14}\,\rm rad\,s^{-1}
\end{equation}
(see~\cite{BOC1970}, Eq.\,(76c) with $S_{1}\simeq0.33MR^{2}\omega_{1}$)
and
\begin{equation}\label{eq:SpinSpinOrbitPrecession}
\begin{split}
\Omega^{orbit}_{S_{1}S_{2}}\simeq\frac{3}{4}0.33\epsilon\bigg(\frac{R}{r}\bigg)^{2}\frac{a^{2}+b^{2}}{r^{2}}\frac{\omega_{1}\omega_{2}}{n}\\
\simeq3.16\cdot10^{-26}\,\rm rad\,s^{-1}
\end{split}
\end{equation}
(see~\cite{BOC1970}, Eq.\,(76d) with $S_{1}\simeq0.33MR^{2}\omega_{1}$ and $S_{2}\simeq\frac{1}{2} m(a^{2}+b^{2})\omega_{2}$ for the test cylinders of GG and a worst case estimate). 

Since UFF/WEP tests are differential experiments,  only the differential precession of the orbits of the test cylinders relative to each other is relevant.   For the orbit precessions (\ref{eq:EinsteinOrbitPrecession}) and (\ref{eq:SpinOrbitPrecession}) the difference  between the two cylinders is due to the fact that they are not exactly at the same distance from the center of mass of the Earth, hence:  
\begin{equation}\label{eq:EinsteinOrbitPrecessionDifferential}
\Delta\Omega^{orbit}_{E}\simeq\frac{5}{2}\Omega_{E}\frac{\Delta r}{r}\simeq1.2\cdot10^{-27}\,\rm rad\,s^{-1}
\end{equation}
and 
\begin{equation}\label{eq:SpinOrbitPrecessionDifferential}
\Delta\Omega^{orbit}_{S_{1}}\simeq3\Omega^{orbit}_{S_{1}}\frac{\Delta r}{r}
\simeq1.8\cdot10^{-29}\,\rm rad\,s^{-1} \ \ \ .
\end{equation}
\begin{equation}\label{eq:SpinSpinOrbitPrecessionDifferential}
\begin{split}
\Delta\Omega^{orbit}_{S_{1}S_{2}}\simeq\frac{3}{4}0.33\epsilon\bigg(\frac{R}{r}\bigg)^{2}\frac{\omega_{1}\omega_{2}}{n} \frac{1}{r^{2}}\big(\Delta(a^{2}+b^{2})\big)\\
\simeq2\cdot10^{-26}\,\rm rad\,s^{-1}
\end{split}
\end{equation}

In addition, it has been shown (\cite{BOC1970}, Eq.(46)) that according to GR the quadrupole mass moment of the primary body gives an additional contribution to  the precession of the orbit of the test mass: 
\begin{equation}\label{eq:Q1GRorbitPrecession}
\Omega^{orbit}_{Q_{1}GR}\simeq\frac{9}{4}\epsilon\,J_{2}^{(1)}
\bigg(\frac{R}{r}\bigg)^{2}n
\simeq1.4\cdot10^{-15}\,\rm rad\,s^{-1} \ \ .
\end{equation}
The corresponding differential precession between the orbits of the test masses, with the center of mass offset as in the case of GG, is: 
\begin{equation}\label{eq:Q1GRorbitPrecessionDifferential}
\Delta\Omega^{orbit}_{Q_{1}GR}\simeq\frac{7}{2}\Omega_{Q_{1}GR}\frac{\Delta r}{r}
\simeq1.17\cdot10^{-30}\,\rm rad\,s^{-1} \ \ \ .
\end{equation}

These results  show that the largest differential precession predicted by General Relativity is $\Delta\Omega^{orbit}_{S_{1}S_{2}}$, due to the proper rotation of the Earth coupling with the proper rotation of the test cylinders. However, the angular velocity of differential precession is extremely small, with a period more than $8$ orders of magnitude longer than the age of the universe.

By comparison with the GR effects computed above it is worth recalling   the Newtonian contributions to orbit precession which are already taken into account in the experiments and their numerical simulations. The largest one is the well known precession due to the quadrupole mass moment of the Earth on any point mass moving on an inclined orbit around it.  This effect  is exploited by both Microscope and GG in order to keep the satellite in a sun-synchronous orbit (with an appropriate choice of  the inclination for the selected altitude). It is given by:
\begin{equation}\label{eq:Q1orbitPrecession}
\Omega^{orbit}_{Q_{1}}\simeq\frac{3}{2}J_{2}^{(1)}
\bigg(\frac{R}{r}\bigg)^{2}n
\simeq1.4\cdot10^{-6}\,\rm rad\,s^{-1} 
\end{equation}
yielding  a much smaller differential precession between the orbits of the test masses:
\begin{equation}\label{eq:Q1orbitPrecessionDifferential}
\Delta\Omega^{orbit}_{Q_{1}}\simeq\frac{7}{2}\Omega_{Q_{1}}\frac{\Delta r}{r}
\simeq1.2\cdot10^{-21}\,\rm rad\,s^{-1}
\end{equation}

It has been shown    in~\cite{BOC1975}, Eq.\,(71) that an additional  Newtonian precession of the orbit occurs if the secondary body has a non zero quadrupole mass moment of its own, coupling with the monopole of the Earth. In this case the dependence on the average size of the test body (squared) makes the effect many orders of magnitude  smaller than the previous one. With the typical numbers of GG we have, for the largest body:
\begin{equation}\label{eq:Q2orbitPrecession}
\Omega^{orbit}_{Q_{2}}\simeq\frac{3}{2}J_{2}^{(2)}
\bigg(\frac{\bar r}{r}\bigg)^{2}n
\simeq6.4\cdot10^{-21}\,\rm rad\,s^{-1}
\end{equation}
where $J_{2}^{(2)}$ is the quadrupole mass moment of the test body  (defined similarly to the quadrupole mass moment $J_{2}^{(1)}$ of the Earth in the expansion of its gravitational field in multipole mass moments) and  $\bar r$  is the average size (half the sum of the inner and outer radius of the test cylinder). The numerical estimate refers to GG (worst case value), and the corresponding differential precession turns out to be only   one order of magnitude smaller: 
\begin{equation}\label{eq:Q2orbitPrecessionDifferential}
\Delta\Omega^{orbit}_{Q_{2}}
\simeq7.5\cdot10^{-22}\,\rm rad\,s^{-1}\  ,
\end{equation}
a value slightly smaller than the differential precession (\ref{eq:Q1orbitPrecessionDifferential}) due to the quadrupole mass moment of the Earth. 

As we can see by comparing  (\ref{eq:EinsteinOrbitPrecession}) and (\ref{eq:Q1orbitPrecession}), the largest precession due to GR is $6$ orders of magnitude smaller than the largest Newtonian precession, while in the case of differential precession the Newtonian ones dominate by $4$ orders of magnitude (see  (\ref{eq:Q1orbitPrecessionDifferential}) and (\ref{eq:Q2orbitPrecessionDifferential}) in comparison  with the largest GR differential precession (\ref{eq:SpinSpinOrbitPrecessionDifferential})). 

In order to assess  how much these orbit precessions affect the  test we use the variation of the elements perturbative equations in the form of Gauss (see~\cite{MNF}), in which the time variation of the orbital elements are expressed in terms of the radial, transverse and out-of-plane component of the perturbing acceleration $a_{R}$, $a_{T}$ and $a_{W}$ respectively. For the effects on the pericenter and the node we use Eqs, (3.43) and (3.47) of~\cite{MNF} and find the following relationships between a (differential) orbit precession rate and the corresponding (differential) components of the perturbing acceleration that generates it:
\begin{equation}\label{eq:aoutoftheplane}
\Delta a_{_{W}}\simeq v\Delta\Omega_{Q_{2}}^{orbit}\simeq5.6\cdot10^{-18}\,\rm m\,s^{-2}
\end{equation}
\begin{equation}\label{eq:aradialandtransverse}
\Delta a_{_{R}}\simeq\Delta a_{_{T}}\simeq e v\Delta\Omega_{Q_{2}}^{orbit}\simeq5.6\cdot10^{-20}\,\rm m\,s^{-2} \ \ .
\end{equation}
As expected, they are related through the orbital velocity $v$, and --for the same precession of the orbit-- the  radial and transverse components of the perturbation involved are smaller than the out-of-plane component by as much as  the orbital eccentricity $e$ (we have used the maximum value required for GG $e\lesssim0.01$;  for Microscope the requirement is $e\lesssim0.005$). 

Only the radial component  $\Delta a_{_{R}}$ would compete with the violation signal. The corresponding spurious contribution to violation comes from the Newtonian differential precession  and amounts to:
\begin{equation}\label{eq:etaorbitprecession}
\eta_{orbit-precession}=\frac{\Delta a_{_{R}}}{g(h)}\simeq
\frac{5.6\cdot10^{-20}}{8.1}
\simeq6.9\cdot10^{-21}
\end{equation}
 which is more than $3$ orders of magnitude smaller than the GG target. As for the contribution from the largest GR effect (\ref{eq:SpinOrbitPrecessionDifferential}), it is $4$ orders of magnitude smaller still.

\section{General relativistic effects on the spin axes of the test masses}
\label{Sec:GReffectsSpnAxes}

In the framework of General Relativity the spin axes of the  test masses are subjected to the following precessions:
\begin{equation}\label{eq:DSprecession}
\Omega^{spinaxis}_{E}\simeq\frac{3}{2}\epsilon n
\simeq10^{-12}\,\rm rad\,s^{-1}
\end{equation}
 caused by the primary body (the Earth) regardless of its rotation (also known as De Sitter precession; see e.g.~\cite{BOC1970}, Eq.\,(42)), and
\begin{equation}\label{eq:SpinPrecession-S1}
\Omega^{spinaxis}_{S_{1}}\simeq\frac{0.33}{2}\epsilon\bigg(\frac{R}{r}\bigg)^{2}\omega_{1}
\simeq6.3\cdot10^{-15}\,\rm rad\,s^{-1}
\end{equation}
due to the proper rotation of the Earth with spin angular momentum $S_{1}\simeq0.33MR^{2}\omega_{1}$ (also known as Lense-Thirring precession; see e.g~\cite{BOC1970} Eq.\,(29)).  The corresponding differential precessions between the spin axes of the two cylinders, due to  the fact that they are not exactly centered on each other, are (GG case):
\begin{equation}\label{eq:DSprecessionDifferential}
\Delta\Omega^{spinaxis}_{E}\simeq\frac{5}{2}\Omega^{spinaxis}_{E}\frac{\Delta r}{r}
\simeq6.2\cdot10^{-28}\,\rm rad\,s^{-1}
\end{equation}
and 
\begin{equation}\label{eq:SpinPrecession-S1Differential}
\Delta\Omega^{spinaxis}_{S_{1}}\simeq3\Omega^{spinaxis}_{S_{1}}\frac{\Delta r}{r}\simeq4.6\cdot10^{-30}\,\rm rad\,s^{-1} \ \ .
\end{equation}

A by far larger precession of the spin axes of the test cylinders is  Newtonian, due to the fact that they have a non zero fractional difference of their principal moments of inertia $\frac{I_{z}-I_{x}}{I_{x}}=\frac{\Delta I}{I_{x}}$, $z$ being the direction of the symmetry/rotation axis. According to~\cite{BOC1975}, Eq.\,(47), the precession rate of the spin axis is: 
\begin{equation}\label{eq:SpinPrecession-Q2Newton}
\Omega^{spinaxis}_{Q_{2}S_{2}}\simeq\frac{1}{2}\frac{\Delta I}{I_{x}}
\frac{n^{2}}{\omega_{2}}\simeq1.3\cdot10^{-9}\,\rm rad\,s^{-1}
\end{equation}
where the numerical estimate refers to the GG test cylinders (worst case: $\frac{\Delta I}{I_{x}}\simeq0.014$). Note that this angular precession rate is proportional to the ratio $\frac{n^{2}}{\omega_{2}}$ between the orbital mean motion $n$  squared and the rotation angular velocity of the test cylinder $\omega_{2}$, thus implying that 
for test cylinders with the same orbital velocity those  which spin faster have a slower precession rate (the ratio of the spin rates  is about  $900$ to $1$  between GG and Microscope). The corresponding differential precession between the test cylinders is not much smaller because they cannot have values of $\frac{\Delta I}{I_{x}}$ exactly (or very nearly) equal. In GG, by requiring  that the difference is of $2\cdot10^{-3}$ the differential precession rate of the spin axes is: 
\begin{equation}\label{eq:SpinPrecession-Q2NewtonDifferential}
\Delta\Omega^{spinaxis}_{Q_{2}S_{2}}\simeq\frac{1}{2}\Delta\Big(\frac{\Delta I}{I_{x}}\Big)
\frac{n^{2}}{\omega_{2}}\simeq1.8\cdot10^{-10}\,\rm rad\,s^{-1}
\end{equation}
with a differential precession period of one thousand years. In the planned one year duration of the mission  the differential precession angle amounts to about  $0.3^{\circ}$ while each spin axis  precesses by about $2.4^{\circ}$. We recall that in GG the test cylinders spin around their symmetry axes, hence the rotation is stable against small perturbations. In addition, the spin frequency  is higher than the normal mode frequencies in the plane perpendicular to the spin/symmetry axis: a dynamical condition which is known to ensure natural damping of the conical modes (precessions) (see~\cite{Genta}, \cite{GGnondragfree}). 

A realistic estimate of spin axes precessions in the case of Microscope would require knowledge of the geometrical and mass properties of the test cylinders  and of the rotation control of the system. However, it is apparent that the Newtonian framework under which the experiment has been designed is fully adequate to the task, since precessions due to General Relativity are many many orders of magnitude smaller than the  Newtonian ones.

\section{Effects on the readout laser gauge}
\label{Sec:LaserGauge}

In GG the relative displacements of the test cylinders in the sensitive plane perpendicular to the spin/symmetry axis are read by a laser interferometry gauge. It has been proposed in 2010  by M. Shao (JPL) in substitution of the originally planned capacitance readout (tested in the laboratory prototype of GG)  because of its numerous advantages~\cite{Shao2010}.
A violation signal at the target level of $10^{-17}$  shows up as a  $0.6\,\rm pm$ displacement between the centers of mass of the test cylinders pointing to (or away from) the center of mass of the  Earth at the spin frequency of $1\,\rm Hz$ (after up-conversion by rotation from the much lower orbital frequency of $1.7\cdot10^{-4}\,\rm Hz$). The laser gauge is expected to have a displacement noise of $\frac{1\,\rm pm}{\sqrt{\rm Hz}}$ at $1\,\rm Hz$.

In a spinning experiment the readout is obviously co-rotating with the system. As reported in Sec.\,\ref{Sec:Introduction}, concerns have been expressed by the SARP panel of ESA, about the effects of rotation  (of the tests cylinders and the Earth)  on the readout laser gauge.

The laser gauge designed for GG is presented in~\cite{LIG2015} where some key sources of noise are discussed on the basis of the results of specific lab tests. In~\cite{LIG2016} a measured displacement noise of $\frac{3\,\rm pm}{\sqrt{\rm Hz}}$ is reported at $1\,\rm Hz$ and the onset of a spurious displacement in the presence of rotation (Sagnac effect) is discussed and quantified.
 The issue is  as follows: if --from  the point of separation to the point  of recombination-- the interfering laser rays happen to enclose a non-zero area, and the axis normal to this area has a component along the the rotation axis, then the laser rays traveling in the sense of rotation and those traveling opposite to it (both at the speed of light) do have different flight times, yielding a spurious  interference signal. In laser gyros  this `spurious' signal is used to measure the rotation angular velocity perpendicular to the area enclosed by the gyro (e.g. the diurnal rotation velocity the Earth).

In the case of GG the laser rays are (nominally) aligned, and the area enclosed  from separation to recombination is zero, hence there should be no spurious displacement. However, a non-zero area arises in the presence of a misalignment. As  shown in~\cite{LIG2016}, a misalignment of $10\,\rm \mu m$  (with a typical $20\,\rm cm$ separation from beam launcher to target) results in a spurious displacement of $\simeq4\cdot10^{-14}\,\rm m$. It is 2 orders of magnitude smaller than the signal; moreover, only its time variation at the spin frequency does compete with the violation signal (a constant bias doesn't matter). There is even a better way out: the lasers can be arranged in such a way  that the 
angular velocity vector of rotation has no component perpendicular to the non-zero area resulting from the misalignment  and therefore there is (nominally) no Sagnac effect. In reality, the effect is reduced even further. This strategy has been followed from the start in the design of the GG laser gauge.
 
 As far as the effects of  the rotation of the Earth on the laser gauge of GG are concerned, it is known that the rotation  of a celestial body (and also its flattening) do  affect the angle of deviation and the propagation time of light rays which, in their journey  from emitter to receiver happen to pass close to the body (see e.g.~\cite{Klioner1991}). In high precision astronomical measurements such as those carried out by the GAIA mission of ESA these effect are indeed  carefully calculated and taken into account in the framework of General Relativity.  However, in  GG there is no measurement over astronomical distances: the laser interferometer works inside the spacecraft, with  optical path differences (in between the targets located, respectively,  on the outer surface of the inner cylinder and on the inner surface of the outer one) of $2\,\rm cm$ or less, and a path length of about $20\,\rm cm$. 
 
 We therefore don't expect that any effect from the spin (and flattening) of the Earth on the laser interferometry readout of GG should be taken into account. 
More importantly, there is  recent experimental evidence which supports this conclusion.  A heterodyne laser gauge similar to the one of GG, though more demanding and complex, has recently flown on LISA pathfinder (LPF)~\cite{LPF2016}. In this case the optical  path difference is  $\simeq38\,\rm  cm$,  which is the separation distance between the test masses  that constitute the mirrors which, in  the final gravitational wave interferometer LISA, will be located in different spacecraft about  $5$ million km away  from each other.   The laser gauge of LPF  has been designed to  achieve   low noise down to  $7\cdot10^{-4}\,\rm Hz$ (while in GG the violation signal is up-converted by rotation to $1\,\rm Hz$). 
Reaching this level of  noise at such low frequencies and over $38\,\rm cm$ separation of the test masses  requires --among other things-- the frequency of the laser to be stabilized. This is a demanding requirement in space, which does not apply in the case of GG because of the $1\,\rm Hz$ frequency of the signal and $2\,\rm cm$ maximum separation.
 The displacement noise as  measured by LPF   above $0.06\,\rm Hz$ is reported to be  of $\frac{0.035\,\rm pm}{\sqrt{\rm Hz}}$ (about $30$ times lower than required for GG at $1\,\rm Hz$); it is interpreted by the authors as entirely due to the interferometer  and  no evidence is reported of any effect due to the rotation and flattening  of the Earth~\cite{LPF2016}. In addition, it turns out to be about $100$ times lower in absence of weight  than obtained in ground tests before launch.

\section{Conclusions}
\label{Sec:Conclusions}
 
 The Microscope satellite, launched on 25 April 2016, is testing the Universality of Free Fall and the Weak Equivalence Principle in the gravitational field of the Earth aiming at two orders of magnitude improvement  over the current best tests. In a similar orbit but with a different experiment design, the GG satellite aims at a test one hundred  times better than Microscope. They must detect extremely small differential accelerations acting between test cylinders of different composition, ruling out any competing effect which is due to known physics.  
 
 In 2015, during the evaluation process of the space mission proposals shortlisted as candidates for the medium size mission M4 of the European Space Agency, the panel appointed by the Agency to evaluate GG made the point that while the  experiment has been designed in the framework of Newtonian physics, it should instead take into account general relativistic effects, in particular those which involve the spin angular momentum of the source body (the Earth) and that of  the test cylinders. 
 
 We have carefully analyzed all known general relativistic effects on the test cylinders of GG and Microscope  showing that they are all negligible by and large.

\textbf{Acknowledgements.} Thanks are due to Norbert Wex, David Lucchesi, Angelo Tartaglia, Slava Turyshev, Juergen Mueller, Alberto Anselmi and Maria Teresa Crosta for their contributions. Support from ESA in the development of a low noise laser gauge over small distances is gratefully acknowledged.

\end{document}